\PassOptionsToPackage{svgnames,dvipsnames}{xcolor}
\documentclass[journal]{vgtc}

\onlineid{0}
\vgtccategory{Research}

\usepackage{graphicx}
\usepackage{xcolor}
\usepackage{mathptmx}
\graphicspath{{figures/}{./}}
\AtBeginDocument{\expandafter\def\expandafter\UrlBreaks\expandafter{\UrlBreaks\do\/\do\*\do\-\do\~\do\'\do\"\do\-}}

\definecolor{bgcolor}{rgb}{0,0.42,0.64}
\definecolor{syellowcolor}{rgb}{0.98,0.64,0.098}
\definecolor{sidebartint}{rgb}{0.93,0.94,1.0}

\newcommand{\chapteri}[1]{#1}

\makeatletter
\newcommand{\nocaptionfloat}{\global\@capstartfalse}
\makeatother

\renewcommand{\manuscriptnotetxt}{}

\title{Agentic Dashboards: A New Design Paradigm for Visual Analytics}

\author{Sam Yu-Te Lee, Yun-Hsin Kuo, and Kwan-Liu Ma}
\authorfooter{%
\item
  All authors are with the University of California, Davis.
  E-mail: \{ytlee, yskuo, klma\}@ucdavis.edu
}

\abstract{%
Visual analytics systems traditionally combine analytical methods and interactive visualizations into a fixed application specified at the design phase. Coding agents offer a flexible approach: experts could conduct analysis through natural-language instructions. We envision an agentic dashboard as a persistent visual analytics workspace where an agent makes decisions about analytical methods, views, and their coordination within a designer-defined decision space. This approach shifts part of the visualization designer's role from engineering a fixed interface to defining the decisions an agent may make at runtime. We illustrate how such a dashboard could operate with current capabilities and propose criteria for evaluating its behavior and potential failures. As analytical workflows become more fluid and agent-mediated, visualization can help experts maintain awareness, coordinate with agents, and exercise judgment.
}


\begin{document}

\firstsection{Introduction}
\maketitle

\chapteri{V}isual analytics (VA) systems are conventionally designed for known application scenarios, analytical goals, and data. Based on application requirements, designers select or develop analytical methods, design interactive views during the design phase, and integrate them into an application with a fixed workflow. Domain experts then carry out analysis within that application.

Agents introduce flexibility to this approach by planning and executing multistep analytical workflows at runtime~\cite{zhao2025lightva}, adapting the workflow as the expert's goals evolve throughout the analysis.
Increasingly, agents are positioned as collaborators that share responsibility for directing the analysis process~\cite{dhanoa2025agentic, lee2026treadstone}.
Agentic visual analytics has emerged as a term for such systems, with a recent survey characterizing agent roles across the visual analytics workflow~\cite{luo2026agenticva}.
These systems tend to demonstrate individual capabilities facilitated by agents, such as adaptability or steerability.

Bringing these capabilities together in a unified environment suggests a new class of visual analytics systems, \textbf{agentic dashboard}, that support diverse application scenarios through runtime adaptation.
This shift also changes the role of visualization designers. Rather than specifying analytical workflows and their associated visualization components during the design phase, designers would define a decision space that guides and constrains the agent's selection of analytical methods, visual representations, and their coordination at runtime.
In light of agents' growing role in data analysis, this article explores this emerging direction by examining how design decisions shift to runtime and illustrating the concept of an agentic dashboard through an application scenario.
We then discuss how designers can bound and evaluate agent behavior. Finally, we consider the broader role  of visualization in this future and the research opportunities.

\section{Automating the Workflow}

To understand how agents change the design of conventional VA systems, we distinguish three layers: an \emph{algorithm} layer of analytical methods, a \emph{control} layer of hyperparameters and workflow, and a \emph{view} layer of visual representations experts inspect, coordinate, and interact with. 
Traditionally, visualization designers integrate all three layers into an application through collaboration with domain experts and iterative design.

Agents loosen this design-phase coupling, starting with the \emph{control} layer. 
Rather than following a predefined workflow, an agent can assemble and execute analytical steps at runtime by invoking existing analysis tools and data resources through API, guided by goals and data specified by the expert in natural language~\cite{stahle2026vacp, lee2026videe}. 

When existing tools are insufficient, a coding agent can implement a missing method and refine the implementation based on execution results and evaluations, extending automation to the \emph{algorithm} layer.
To illustrate this possibility, we translated GreenMine~\cite{lee2025greenmine}, a VA system designed to support taxonomy construction for environmental scientists, into a collection of textual specifications. A coding agent can follow these specifications to reproduce the intended analysis workflow in GreenMine, while determining the specific implementations of tasks such as transcript segmentation and code assignment at runtime.

Automation of the \emph{view} layer remains less mature, although research has explored generating coordinated, interactive views~\cite{pandey2023medley,chen2025interchat}. In current practice, an agent may produce reports or isolated visualizations without linked interactions, requiring experts to translate visual selections into natural-language instructions. Across turns of interaction with agents, agent outputs can become scattered across the conversation and detached from the initial assumptions and provenance. Warnings and checks may similarly lose their context. 
This lack of continuity makes it difficult for experts to interpret and validate agent outputs, potentially pushing experts out of the loop and increasing the risk of overreliance~\cite{mitchell2026agents}.

To address this limitation, we next examine how an \emph{agentic dashboard}, a persistent visual analytics workspace, could keep analysis results connected to their provenance while supporting expert steering.

\begin{figure}[!t]
\centering
{\setlength{\fboxsep}{0pt}%
\setlength{\fboxrule}{0.4pt}%
\fcolorbox{black}{sidebartint}{%
\begin{minipage}{\dimexpr\columnwidth-2\fboxrule\relax}
\colorbox{bgcolor}{%
\begin{minipage}{\linewidth}
\vspace{6pt}
\hspace{8pt}\begin{minipage}{\dimexpr\linewidth-16pt\relax}
{\color{white}\bfseries\scriptsize ILLUSTRATIVE EXAMPLE}\par
\vspace{2pt}
{\color{white}\bfseries\large GreenMine: From Workflow\\[-1pt]to Agent Skills}
\end{minipage}
\par\vspace{6pt}
\end{minipage}}
\begin{minipage}{\linewidth}
\vspace{8pt}
\small
\raggedright
\setlength{\leftskip}{8pt}
\setlength{\rightskip}{8pt plus 1fil}
\setlength{\parindent}{0pt}
\hyphenpenalty=10000
\exhyphenpenalty=10000

{\color{bgcolor}\bfseries\footnotesize WHAT GREENMINE SUPPORTS}\par
\vspace{3pt}
Built specifically around the DPSIR framework, GreenMine lets domain experts construct and apply a taxonomy to interview transcripts, refine codes and themes, and inspect uncertainty and relationships without writing code \cite{lee2025greenmine}.

\vspace{6pt}
\begin{center}
\setlength{\fboxsep}{3pt}%
\renewcommand{\arraystretch}{0.85}%
\begin{tabular}{@{}c@{\hspace{3pt}}c@{\hspace{3pt}}c@{}}
\fcolorbox{bgcolor}{white}{\parbox{0.34\linewidth}{\centering\footnotesize\bfseries Build\\DPSIR taxonomy}}
& {\color{syellowcolor}\bfseries$\longrightarrow$} &
\fcolorbox{bgcolor}{white}{\parbox{0.34\linewidth}{\centering\footnotesize\bfseries Apply to\\transcripts}}\\
&& {\color{syellowcolor}\bfseries$\downarrow$}\\
\fcolorbox{bgcolor}{white}{\parbox{0.34\linewidth}{\centering\footnotesize\bfseries Inspect\\results}}
& {\color{syellowcolor}\bfseries$\longleftarrow$} &
\fcolorbox{bgcolor}{white}{\parbox{0.34\linewidth}{\centering\footnotesize\bfseries Refine codes\\and themes}}
\end{tabular}
\end{center}

\raggedright
\setlength{\leftskip}{8pt}
\setlength{\rightskip}{8pt plus 1fil}

\vspace{4pt}
{\color{bgcolor}\bfseries\footnotesize THE WORKFLOW AS SKILLS}\par
\vspace{4pt}
\noindent\includegraphics[width=\dimexpr\linewidth-16pt\relax]{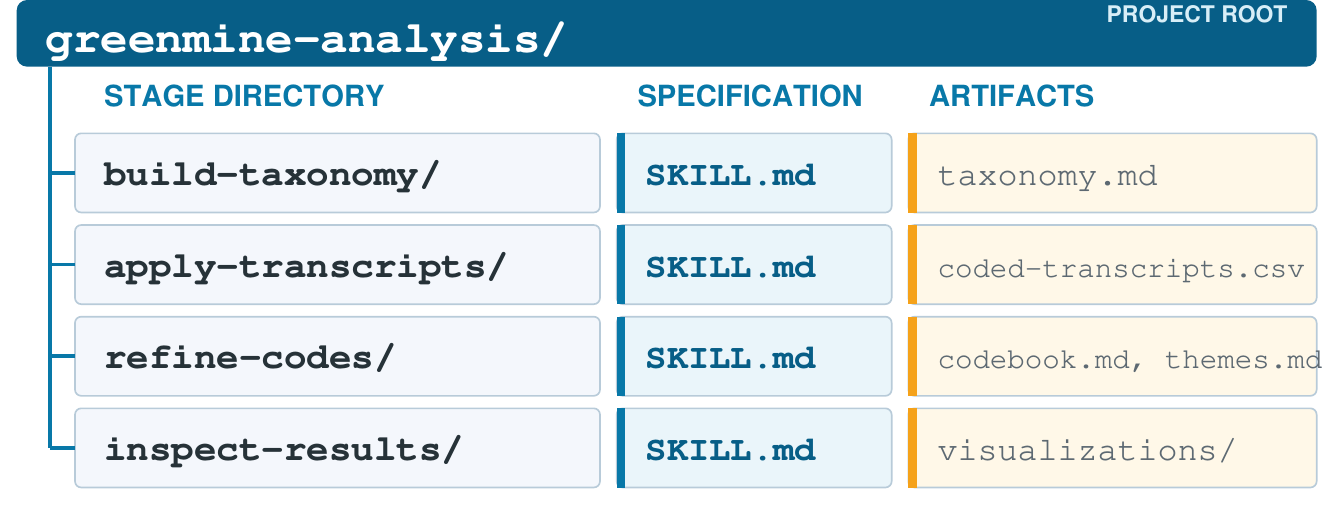}\par

\vspace{6pt}
{\setlength{\fboxsep}{5pt}%
\colorbox{white}{%
\parbox{\dimexpr\linewidth-16pt-2\fboxsep\relax}{%
{\color{bgcolor}\bfseries\footnotesize TAKEAWAY}\par
\vspace{2pt}
\footnotesize The analytical stages and domain constraints remain, while the coding agent determines how to implement and connect them at runtime.}}}

\vspace{8pt}
\end{minipage}
\end{minipage}}}
\nocaptionfloat
\end{figure}

\begin{figure*}[t]
\centering
\includegraphics[width=\textwidth]{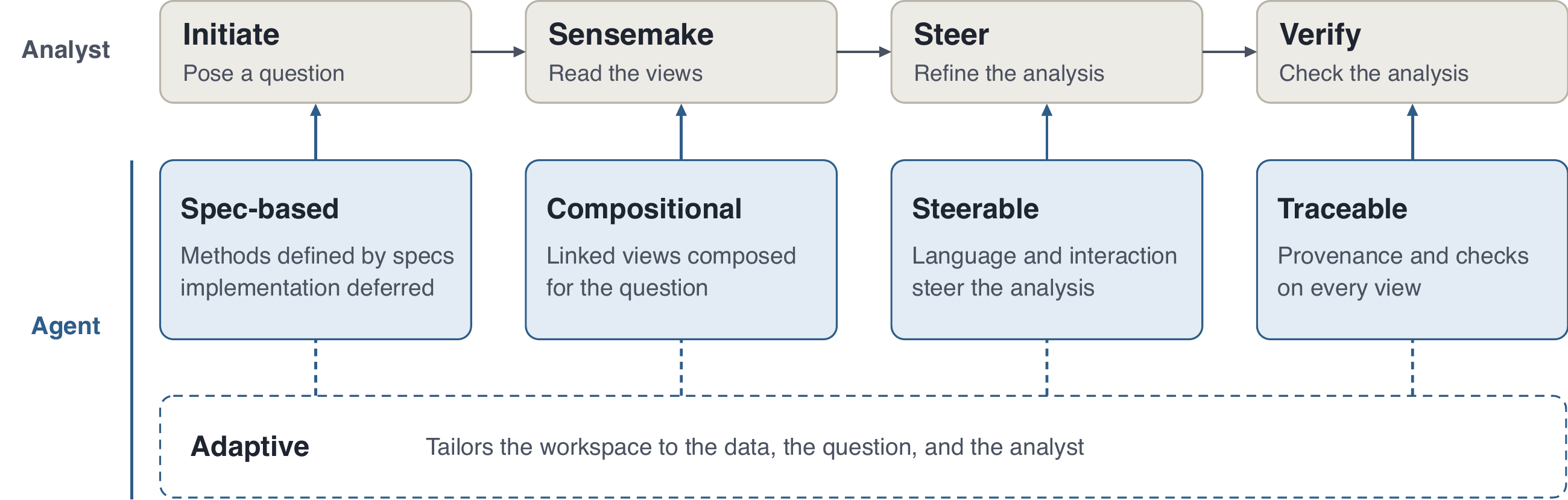}
\caption{Five properties distinguish an agentic dashboard from a sequence of generated outputs. Each of the four agent-side properties supports the expert activity above it, while Adaptive spans all four expert activities.}
\label{fig:properties}
\end{figure*}

\section{Agentic Dashboards}
An \emph{agentic dashboard} delegates some decisions about analytical methods, views, and their coordination to an agent at runtime, within a designer-defined decision space. This delegation does not imply unconstrained autonomy: designers define the boundaries of agent decisions, while the workspace provides experts with mechanisms to inspect and steer agent decisions \cite{heer2019agency}. 
Beyond delegation, an agentic dashboard provides a persistent workspace in which views, their coordination, and analytical states are maintained across interactions.
This continuity allows experts to revisit results in a stable environment that preserves the connections to the data and interactions.
Drawing from existing literature on agentic visual analytics~\cite{dhanoa2025agentic}, we characterize how such a workspace operates through five desirable properties.

\noindent\textbf{Spec-based.} An analysis is defined through a textual specification describing its goals, application context, data input, desired outputs, interactivity, constraints, and validation checks, as detailed in the following section. 
This separates what the analysis should accomplish from how it is implemented, allowing implementation decisions to be deferred until runtime. The specification also captures designers' understanding of experts' requirements and preferences, guiding runtime decisions toward the intended analytical purpose rather than generic solutions.

\noindent\textbf{Compositional.} The agent assembles analytical methods and views into a coherent workspace at runtime. Composition concerns not only which views are presented and how they are arranged, but also how shared analytical state is maintained and propagated among them. 
This coordination allows experts to compare complementary evidence without having to reconstruct the analytical context across separate outputs.

\noindent\textbf{Steerable.} Experts can influence the agent's decisions through natural language and direct interaction with the workspace. Visual interactions such as selecting and brushing data can provide context for redirecting the ongoing analysis. Experts can thus refine the delegated analysis within its visual context without having to translate selections or observed patterns into words.

\noindent\textbf{Traceable.} The methods, data subsets, parameters, assumptions, implementations, and validation checks behind each result remain available for inspection. 
These records remain accessible alongside the corresponding views, allowing experts to inspect how each result was produced without searching through a conversational transcript.
Without such records, a plausible visualization may conceal inappropriate data, methods, or assumptions, potentially leading to misinformed decisions.
Traceability, however, does not establish validity: an inspectable analytical process does not guarantee correct results.

\noindent\textbf{Adaptive.} The agent can revise analytical workflows and reconfigure the workspace as the expert's questions, data, or analytical state change. These adjustments build on the existing workspace while preserving relevant analytical context, including prior selections and views. Adaptation allows the analysis to evolve with both the expert's analysis goals and working preferences rather than following a fixed sequence. 

Recent agentic VA systems have begun to explore aspects of these properties~\cite{luo2026agenticva}. 
An agentic dashboard brings them together in a persistent workspace, where agent decisions remain bounded by a designer-defined decision space and analytical results remain traceable across interactions. 
Next, we illustrate how such a system could support analysis in practice. 

\begin{figure*}[t]
\centering
\includegraphics[width=\textwidth]{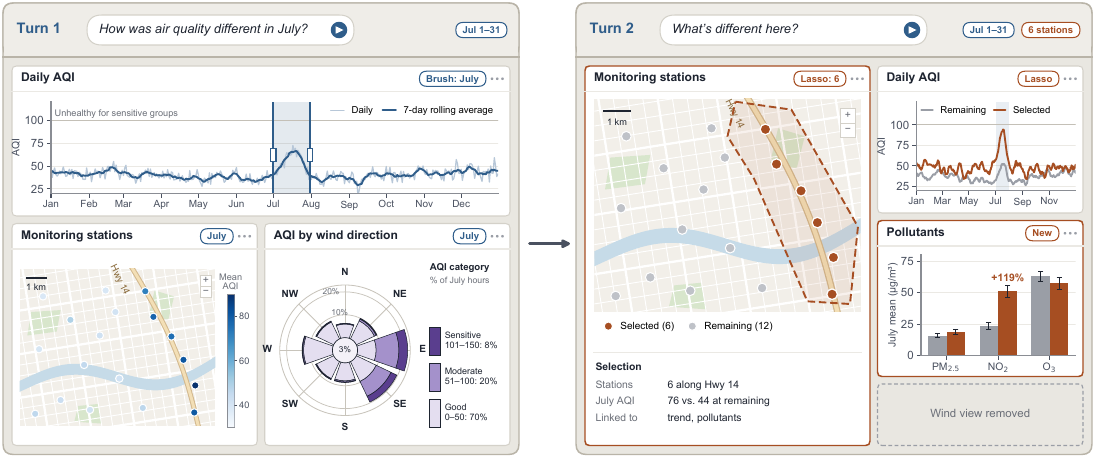}
\caption{An agentic dashboard across two interaction turns. A question about July composes three linked views; a lasso and follow-up question then recompose the workspace around selected highway stations while preserving the analytical state.}
\label{fig:example}
\end{figure*}
\section{The Decision Space}

During the design phase, designers define a specification that guides and bounds the decisions an agent can make at runtime. We organize the specification into three categories, providing a starting point for future work to extend.

\noindent\textbf{Purpose.} This category includes \emph{analysis goals} and \emph{application context}. Analysis goals describe what experts need to understand or decide, guiding agents to select suitable analysis methods~\cite{wu2025requirement}. Application context identifies the intended experts and use setting, informing the agent's choice of terminology and visual representations. 
We consider analysis goals and application context required elements of the decision space, since goals establish what the analysis should accomplish, while context informs what approaches are appropriate for the intended use.

\noindent\textbf{Inputs, results, and use.} This category includes \emph{data input}, \emph{desired outputs}, and \emph{interactivity}. 
Data input identifies relevant data resources and specifies metadata or contextual information that the agent should not infer on its own.
Desired outputs specify evidence types or coordinated views, while leaving implementation and layout decisions to the agent. 
Interactivity defines what user interactions should affect analysis and how they propagate across views.
While agents could infer or determine these elements at runtime, explicitly specifying them allows designers to shape the expert's analytical experience, particularly what the workspace presents and how experts interact with it.

\noindent\textbf{Bounds and assurance.} This category contains \emph{constraints} and \emph{validation checks}. Constraints define permissions or approval conditions that limit the agent's decisions. Validation checks specify tests and responses to failures, such as issuing a warning when a reliability threshold is not met.
We consider both elements essential to the decision space: constraints determine what the agent is permitted to do, while validation checks assess whether agent decisions and analysis results meet the specified analysis goals. 

Next, we illustrate how an agent operates within this specification across interactions.

\section{Illustrative Example}
Consider an expert investigating a city's air-quality data, including daily air-quality index (AQI) measurements from monitoring stations, wind direction, and pollutant levels. The expert begins by asking, ``How was air quality different in July?'' Guided by an AQI analysis specification written by the dashboard's designers, the agent composes three views: a daily AQI trend, a station map, and AQI grouped by wind direction (Figure~\ref{fig:example}). It also links a July brush in the AQI trend to the other two views, allowing the selected period to propagate across the workspace.

The expert then selects stations along a highway in the map and asks, ``What is different here?'' Together, the selection and question indicate which stations to examine and what comparison to make. In response, the agent enlarges the map, separates the AQI trend into selected and remaining stations, adds a pollutant comparison that highlights NO$_2$, and proposes removing the now less relevant wind view, which the expert approves. The station selection remains linked as the workspace adapts to the expert's evolving question.

Each view remains linked to records of its data subset, analytical method, assumptions, and validation checks. The expert can therefore inspect why NO$_2$ was highlighted and assess whether the observed difference is reliable. 
Together, these interactions illustrate how experts can steer the analysis, agents can adapt the workspace and its coordination, and retained provenance can support verification.

\section{Comparing System Paradigms}

An agentic dashboard sits between a predesigned VA system and a general-purpose coding-agent environment in the extent to which design decisions are specified during the design phase or made at runtime. 
Each balances predictability and flexibility differently, making them suitable for different analysis scenarios.

A predesigned VA system is most appropriate for recurring analyses with stable requirements and high reliability demands. Its analytical workflows can be validated and optimized during the design phase, supporting predictable and reliable behavior at runtime. A stable interface also allows experts to develop familiarity through repeated use, improving efficiency in their day-to-day analysis.
Unanticipated questions may require redesign or another environment, but this limitation is acceptable when analyses follow established workflows and reliability takes priority.

A coding-agent environment is better suited to open-ended or one-off investigations involving questions that are difficult to anticipate. 
An agent can assemble analysis workflows from existing methods at runtime, allowing experts to pursue questions without waiting for an application to be developed. However, as the analysis proceeds, experts may need to refine their requests or ask follow-up questions, leaving the results scattered across the conversation and making the analysis difficult to revisit~\cite{Gu2026find_that_chart}.
This limitation may be acceptable for short-lived investigations.


\textbf{We expect agentic dashboards to be most useful for analyses that cannot be fully planned but benefit from a persistent, coordinated workspace}. Compared with a coding-agent environment, an agentic dashboard organizes runtime results into coordinated views that share analytical state, keeping methods and assumptions alongside their corresponding results. 
Compared with a predesigned system, an agentic dashboard could offer a personalized experience by adapting the workspace to the expert’s expertise, working preferences, or preferred visual representations. 
These benefits, however, introduce new design challenges. Designers must carefully construct the decision space, particularly the boundaries of agent autonomy.
In the next section, we discuss general design principles for agentic dashboards.

\section{Relocating Design Decisions}

The scope and quality of analyses supported by an agentic dashboard depend on the decision space designers construct.
Building on the three categories in the decision space introduced earlier, we discuss three considerations for designing the decision space.

\textbf{Designers must make appropriate runtime decisions possible.} They must equip the agent with capabilities and guidance needed to construct analysis workflows that meet experts' requirements. Through collaboration with domain experts, designers can translate analysis requirements and preferences into textual specifications. 
Designers can also provide analysis methods and visualization tools that the agent can use at runtime, along with rules for how these tools can be combined into a coordinated workspace.

\textbf{Designers must make inappropriate decisions difficult or impossible.} Textual specifications alone may be insufficient to enforce constraints that should not depend on the agent's interpretation. Programmatic rules may be necessary to prevent invalid analytical operations and enforce mandatory validation checks. Designers must also define how the agent responds to uncertainty, such as requesting clarification about ambiguous or incomplete information and declining requests outside the decision space.

\textbf{Designers must make consequential decisions visible and reversible.} Decisions that substantially transform data, change analytical methods or visual representations, or discard prior analytical state may require preview or approval. The workspace should preserve sufficient information for experts to inspect these decisions and revise or reverse them when necessary. Designers should establish boundaries for agent autonomy with domain experts, considering applicable standards and potential consequences.

Together, these considerations highlight the challenge of defining a decision space that enables flexible analysis while keeping agent decisions appropriately constrained, inspectable, and reversible. How specifications should vary across domains and levels of risk remains an open question for future research.

\section{Evaluating Agentic Dashboards}
Delegating decisions about analytical methods, views, and their coordination to an agent at runtime introduces new failure modes. The agent may misinterpret the expert's question, select an analysis method that is inappropriate for the data, or construct a misleading visualization. Views may become inconsistent when filters, selections, or analytical results are not updated across the workspace. The agent may also make decisions outside its permitted scope or change the workspace without the expert noticing. These failures are concerning because the dashboard may appear correct even when the underlying analysis is flawed.

Because an agentic dashboard evolves across interactions, the dashboard evolution---not just its current state---should be the unit of evaluation. Evaluations should examine how agent runtime decisions affect the analytical state and subsequent results. They should test whether the agent handles clear, ambiguous, or out-of-scope requests as specified, and whether revisions preserve a valid analytical state across interactions. 
The traceability record should document which specifications, tools, and programmatic rules shaped each decision, allowing evaluators to assess whether the agent operated within its intended decision space.

Evaluations should also examine whether experts can recognize when an analysis goes wrong and recover from consequential errors or unexpected changes. 
The workspace should help experts identify the responsible decisions and their downstream effects, and restore a valid analytical state without losing unaffected work. 
Recovery effort matters: reversibility provides little practical value if experts must reconstruct the entire analysis themselves.


\section{A Broader Role for Visualization}
If agentic dashboards become common, visualization could take on broader roles. We consider two possibilities: supporting the verification of agent-produced findings and coordinating analysis across participants.

As agents take on more analytical work, experts may spend more effort evaluating the findings agents produce. 
Conventionally, visualization helps experts explore data, develop hypotheses, and test their interpretations. 
In agentic dashboards, visualization may support experts in verifying agent findings. This calls for visualizations that expose supporting and contradictory evidence, reveal limitations of the underlying data, and help experts assess whether the findings are justified.
Existing visualization design principles may need to evolve to support this role, but what should change and how remains an open question.

In collaborative analysis, participants may work with agents pursuing different questions or operating under different specifications and methods. A shared agentic dashboard can provide a common reference without requiring everyone to follow the same analytical workflow. By presenting alternative analyses together, the dashboard could reveal differences in interpretation and help participants decide what to investigate next. Visualization could therefore support negotiation as participants work toward shared findings.

However, combining findings across participants can be challenging as agents may merge and reformulate individual results. 
This can lead to \emph{contribution dissolution} \cite{imteyaz2026nobody}, where the originality of individual contributions and responsibility for the merged result become difficult to determine. Although provenance records can document how a result was produced, they cannot establish whether participants agree with the merged interpretation and accept responsibility.
Participants may need opportunities to inspect, question, and approve shared findings. 
Visualization could support this process by presenting an agent's proposed result alongside the contributions it combines. How to design such views and whether they improve accountability remain open questions.

Together, these possibilities suggest roles for visualization beyond automatic chart generation. 
How can a team reconcile findings produced through different agent-mediated workflows? How can visualizations help experts identify agent errors? Addressing these questions would extend the agentic dashboard from an interface for individual analysis into a shared workspace for verifying and coordinating analytical work across participants.

\section{Implications for Visual Analytics Research}
As agents take on more runtime design decisions, we may need to reconsider how VA research contributions are evaluated. Although predesigned systems remain valuable for stable analyses, implementing a complete application from existing tools should not, by itself, establish a novel research contribution. We argue that greater emphasis should be placed on the domain-specific design knowledge and empirical evidence that make these workflows useful and trustworthy.

One opportunity lies in developing and evaluating domain specifications that guide agent runtime decisions. 
Producing these domain specifications requires visual analytics researchers to work closely with domain experts and combine the findings with state-of-the-art analysis techniques, visualization design principles, and analysis task taxonomies. However, existing guidance may not remain appropriate in a particular domain or as analyses evolve across interactions. Contributions on domain specifications should therefore demonstrate that they hold in multi-turn, agent-mediated analysis.

Claimed contributions of an agentic dashboard also need to shift their emphasis to the decision space, particularly on how it supports experts reshaping the dashboard with agents.
With limited visualization literacy and data literacy, we should not expect experts to always make appropriate design and analysis decisions without guidance, and research should examine ways for agents to assist, such as flagging unsuitable views or suggesting alternatives. 
Contributions regarding the decision space should not only be based on success stories, but also demonstrate the effectiveness of guidance and guardrails. 

Ultimately, \textbf{the enduring value of VA research lies in helping domain experts exercise informed judgement over agent-mediated analyses} \cite{elmqvist2025automating}. Whether through predesigned systems or agentic dashboards, visual analytics should help experts generate and verify findings, reconcile different interpretations, and make informed domain decisions.

\bibliographystyle{abbrv-doi-hyperref}
\bibliography{references}

\end{document}